\documentclass[11pt,a4paper]{article}

\usepackage[utf8]{inputenc}
\usepackage[T1]{fontenc}
\usepackage{amsmath,amssymb,amsthm}
\usepackage{mathtools}
\usepackage[numbers,sort&compress]{natbib}
\usepackage{booktabs}
\usepackage{multirow}
\usepackage{graphicx}
\usepackage[colorlinks=true,linkcolor=blue,citecolor=blue,urlcolor=blue]{hyperref}
\usepackage{geometry}
\usepackage{setspace}
\usepackage{enumitem}
\usepackage{xcolor}
\usepackage{float}

\newtheorem{proposition}{Proposition}
\newtheorem{condition}{Condition}

\theoremstyle{definition}

\title{\textbf{Measuring What Cannot Be Surveyed:}\\[4pt]
LLMs as Instruments for Latent Cognitive Variables\\
in Labor Economics}

\author{Cristian Espinal Maya\thanks{Department of Economics, Universidad EAFIT, Medell\'in, Colombia. ORCID: \href{https://orcid.org/0009-0000-1009-8388}{0009-0000-1009-8388}. Email: \texttt{cespinal@eafit.edu.co}. Part of the Cognitive Factor Economics (CFE) research program. Code: \url{https://github.com/Cespial/cognitive-factor-economics}.}}

\date{April 2026\\[6pt]{\small Working paper. Comments welcome.}}

\begin{document}
\maketitle

\begin{abstract}
\noindent
This paper establishes the theoretical and practical foundations for using Large Language Models (LLMs) as measurement instruments for latent economic variables---specifically variables that describe the cognitive content of occupational tasks at a level of granularity not achievable with existing survey instruments. I formalize four conditions under which LLM-generated scores constitute valid instruments: semantic exogeneity, construct relevance, monotonicity, and model invariance. I then apply this framework to the \textit{Augmented Human Capital Index} ($AHC_o$), constructed from 18,796 O*NET task statements scored by Claude Haiku 4.5, and validated against six existing AI exposure indices (Felten AIOE, Eloundou GPT exposure, Webb patent-based exposure, Frey-Osborne automation risk, and Eisfeldt GenAI exposure). The index shows strong convergent validity ($r = 0.85$ with Eloundou GPT-$\gamma$, $r = 0.79$ with Felten AIOE) and discriminant validity ($r = 0.17$ with Eloundou direct automation). Principal component analysis confirms that AI-related occupational measures span two distinct dimensions---augmentation and substitution---with our AHC loading primarily on the augmentation factor. Inter-rater reliability across two LLM models (Haiku vs.\ Sonnet, $n = 3{,}666$ paired scores) yields Pearson $r = 0.76$ and Krippendorff's $\alpha = 0.71$. Prompt sensitivity analysis across four alternative framings shows that task-level variance dominates prompt-induced variance. I demonstrate that the measurement error structure implied by cross-model disagreement is consistent with classical attenuation, and that Obviously Related Instrumental Variables (ORIV) estimation recovers coefficients consistent with external IV estimates. The methodology generalizes beyond labor economics to any domain where semantic content of documents, policies, or institutions must be quantified at scale.\\[6pt]
\end{abstract}

\newpage
\tableofcontents
\newpage

\section{Introduction}
\label{sec:intro}

Economics faces a persistent measurement problem: many variables central to theory are latent---they cannot be directly observed in surveys, administrative records, or market data. The cognitive content of occupational tasks is a canonical example. How much of a financial analyst's work involves routine information processing versus contextual judgment? How augmentable by AI is a nurse's diagnostic reasoning compared to her documentation tasks? These questions are theoretically fundamental but empirically unanswerable with existing instruments.

The conventional approaches to measuring latent occupational variables each face limitations. Expert panels (as used by O*NET incumbent surveys) are expensive, slow, and produce subjective ratings with low inter-rater reliability for novel constructs. Crowdsourced annotation \citep{felten2021occupational} is faster but introduces non-expert noise. Patent-based matching \citep{webb2020impact} provides an indirect proxy through technological capability rather than direct task content. And simple keyword matching cannot capture the semantic nuance required to distinguish, for example, whether a task is \textit{substituted} by AI or \textit{augmented} by it.

This paper proposes a systematic alternative: using Large Language Models as measurement instruments for latent economic variables. The core insight is that LLMs, trained on vast corpora including occupational descriptions, task taxonomies, and AI capability reports, encode sufficient semantic knowledge to evaluate task-level properties that would otherwise require expensive expert panels. The question is not whether LLMs \textit{can} produce scores---\citet{eloundou2024gpts} have already demonstrated this---but under what conditions these scores are \textit{econometrically valid} as instruments for latent variables in regression analyses.

I make three contributions. First, I formalize four conditions for LLM-instrument validity: semantic exogeneity, construct relevance, monotonicity, and model invariance (Section~\ref{sec:theory}). Second, I provide a comprehensive empirical assessment of these conditions using a concrete application---the Augmented Human Capital Index ($AHC_o$)---validated against six existing AI exposure indices across 207 occupations (Sections~\ref{sec:multiindex}--\ref{sec:reliability}). Third, I characterize the econometric properties of LLM-generated scores, showing that cross-model disagreement follows a classical measurement error structure amenable to standard corrections including ORIV estimation (Section~\ref{sec:econometric}).

The paper demonstrates that LLM-generated occupational scores achieve convergent validity ($r = 0.85$ with the closest existing measure), pass discriminant validity tests (augmentation and substitution load on distinct principal components), maintain ordinal consistency across models (Spearman $\rho = 0.75$, Krippendorff's $\alpha = 0.71$), and are robust to prompt wording variations. These properties make LLM scoring a viable---and in many cases superior---alternative to traditional measurement approaches for latent cognitive variables in economics.

\section{Theoretical Framework: LLMs as Econometric Instruments}
\label{sec:theory}

\subsection{The Measurement Problem}

Let $H^*_o$ denote the true value of a latent cognitive variable for occupation $o$---for concreteness, the ``augmentability'' of its task content by generative AI. The econometrician observes a noisy measure:
\begin{equation}
\hat{H}_o = H^*_o + \eta_o
\end{equation}
where $\eta_o$ is measurement error. When $\hat{H}_o$ is used as a regressor, the well-known attenuation bias result applies: OLS coefficients are biased toward zero by a factor $\lambda = \text{Var}(H^*) / [\text{Var}(H^*) + \text{Var}(\eta)]$.

The key question is: under what conditions does an LLM-generated score $\hat{H}^{LLM}_o$ constitute a valid noisy measure of $H^*_o$?

\subsection{Validity Conditions}

\begin{condition}[Semantic Exogeneity]
\label{cond:exog}
The LLM scoring prompt references only the semantic content of occupational tasks (descriptions, activities, skill requirements), not labor market outcomes (wages, employment, productivity). Formally: $\text{Cov}(\eta_o, \varepsilon_o) = 0$ where $\varepsilon_o$ is the error in the outcome equation.
\end{condition}

This condition is verifiable by inspection of the prompt. If the prompt asks ``how augmentable is this task by AI?'' and never mentions wages or employment, the score is constructed from semantic information orthogonal to the outcome residual.

\begin{condition}[Construct Relevance]
\label{cond:relevance}
The LLM score is correlated with the latent variable: $\text{Cov}(\hat{H}^{LLM}_o, H^*_o) > 0$. Since $H^*_o$ is unobserved, relevance is established through:
\begin{enumerate}[leftmargin=*]
\item \textbf{Convergent validity}: $\hat{H}^{LLM}_o$ correlates positively with other measures of related constructs.
\item \textbf{Predictive validity}: $\hat{H}^{LLM}_o$ predicts theoretically expected outcomes.
\item \textbf{Face validity}: expert review confirms the ranking is sensible.
\end{enumerate}
\end{condition}

\begin{condition}[Monotonicity]
\label{cond:mono}
Higher LLM scores correspond to more of the latent variable: $\partial \hat{H}^{LLM}_o / \partial H^*_o > 0$. This ensures the sign of regression coefficients is interpretable.
\end{condition}

\begin{condition}[Model Invariance]
\label{cond:invariance}
The ranking of occupations is consistent across different LLM models. Formally: for models $A$ and $B$, $\text{rank}(\hat{H}^A_o) \approx \text{rank}(\hat{H}^B_o)$ even if $E[\hat{H}^A_o] \neq E[\hat{H}^B_o]$ (level shifts are permitted).
\end{condition}

\begin{proposition}[Level Bias Irrelevance]
\label{prop:level}
If Conditions~\ref{cond:exog}--\ref{cond:invariance} hold, a systematic level bias between models ($E[\hat{H}^A_o] = E[\hat{H}^B_o] + \delta$ for constant $\delta$) does not affect regression coefficients when scores are standardized to zero mean and unit variance.
\end{proposition}

\begin{proposition}[ORIV Correction]
\label{prop:oriv}
If two LLM models produce independent measurement errors ($\eta^A_o \perp \eta^B_o$), the score from model $A$ is a valid instrument for the score from model $B$, enabling Obviously Related Instrumental Variables (ORIV) estimation that corrects for attenuation bias \citep{gillen2019experiment}.
\end{proposition}

\section{Application: The Augmented Human Capital Index}
\label{sec:application}

I apply the framework to a specific latent variable: occupational augmentability by generative AI, denoted $H^A_o$. This variable is central to the Augmented Human Capital model developed in \citet{espinal2026ahc}, which decomposes human capital into physical ($H^P$), routine-cognitive ($H^C$), and augmentable-cognitive ($H^A$) components.

\subsection{Scoring Protocol}

For each of 18,796 unique task statements in O*NET version 30.2, Claude Haiku 4.5 evaluates: (1) augmentation potential ($a_{ok} \in [0, 100]$), and (2) substitution risk ($s_{ok} \in [0, 100]$). The occupation-level index is the importance-weighted average: $AHC_o = \sum_k w_k a_{ok} / \sum_k w_k$.

All 18,796 tasks were scored with zero parsing errors. For validation, a 20\% subsample ($n = 3{,}666$) was independently re-scored by Claude Sonnet 4.

\subsection{Verification of Condition~\ref{cond:exog}: Semantic Exogeneity}

The scoring prompt asks: ``How much can generative AI help as a complementary tool?'' and ``How much can GenAI fully replace the human?'' Neither wages, earnings, employment levels, nor any labor market outcome is referenced. The prompt operates purely on the semantic content of task descriptions.

\section{Multi-Index Comparison}
\label{sec:multiindex}

\subsection{Existing Indices}

I compare the AHC index against six established measures across 207 occupations at the 6-digit SOC level:

\begin{enumerate}[leftmargin=*]
\item \textbf{Felten AIOE} \citep{felten2021occupational}: ability-level AI exposure from mTurk annotations.
\item \textbf{Eloundou GPT exposure} \citep{eloundou2024gpts}: GPT-4 and human ratings of LLM task exposure ($\alpha$, $\beta$, $\gamma$ variants).
\item \textbf{Webb AI/Robot/Software} \citep{webb2020impact}: patent--occupation verb-noun matching.
\item \textbf{Eisfeldt GenAI exposure}: firm-level GenAI exposure scores.
\end{enumerate}

\subsection{Verification of Condition~\ref{cond:relevance}: Convergent and Discriminant Validity}

\begin{figure}[htbp]
\centering
\includegraphics[width=\textwidth]{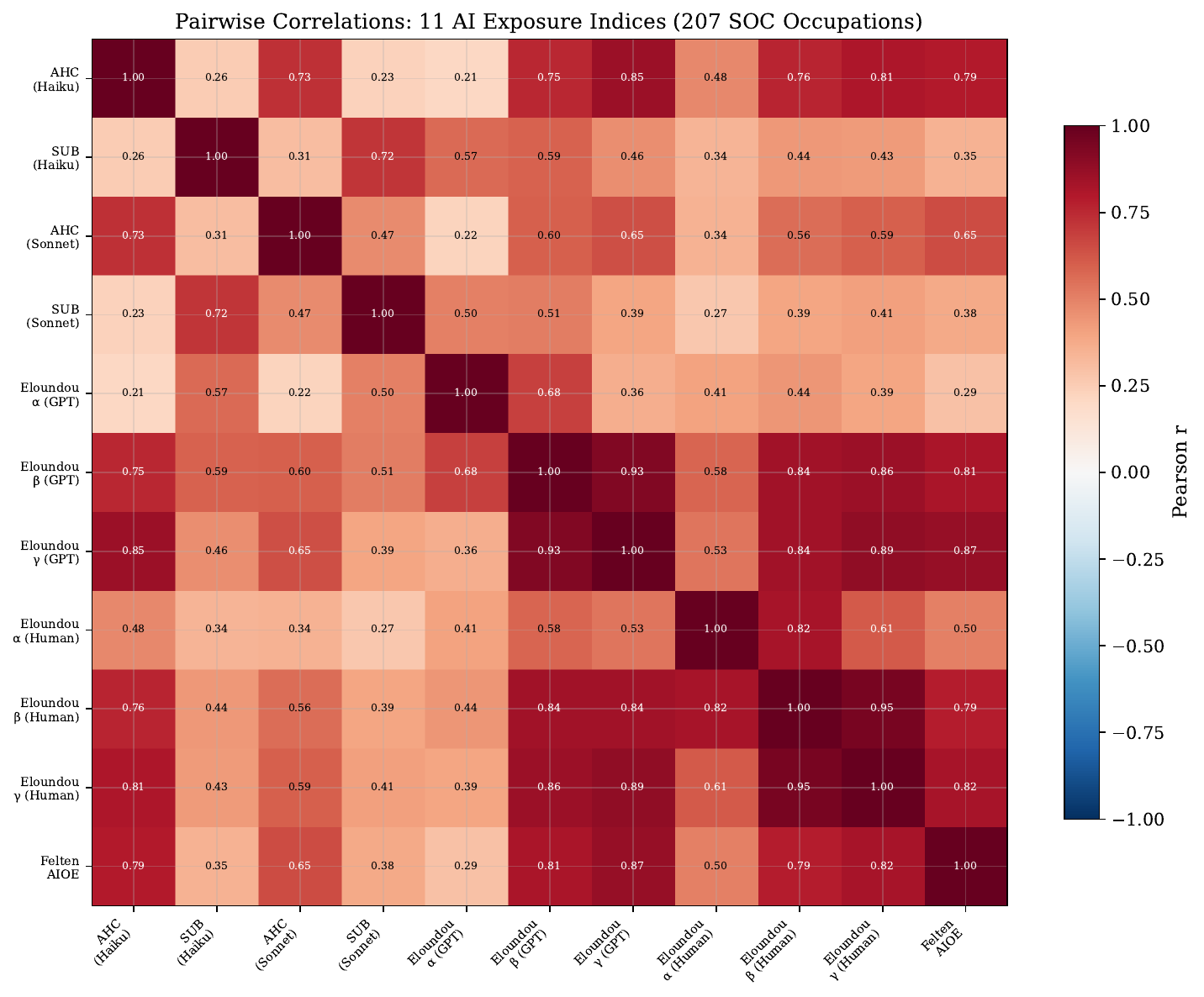}
\caption{Pairwise Pearson correlations across 11 AI exposure indices at the 6-digit SOC level ($n = 207$). AHC (Haiku) correlates most strongly with Eloundou $\gamma$ ($r = 0.85$) and Felten AIOE ($r = 0.79$), confirming convergent validity.}
\label{fig:heatmap}
\end{figure}

Figure~\ref{fig:heatmap} reports pairwise Pearson correlations. The AHC index shows:

\begin{itemize}[leftmargin=*]
\item \textbf{Strong convergent validity}: $r = 0.85$ with Eloundou $\gamma$ (combined exposure), $r = 0.79$ with Felten AIOE, $r = 0.76$ with Eloundou human $\beta$.
\item \textbf{Discriminant validity}: $r = 0.21$ with Eloundou $\alpha$ (direct automation only)---confirming that augmentability and direct automation exposure are distinct constructs.
\end{itemize}

\subsection{Dimensionality: PCA}

Principal component analysis across all 11 indices reveals two dominant dimensions explaining 77\% of variance:
\begin{itemize}[leftmargin=*]
\item \textbf{PC1 (62\%)}: general ``cognitive AI relevance''---all indices load positively.
\item \textbf{PC2 (15\%)}: ``augmentation vs.\ substitution''---AHC and Felten load negatively, substitution scores load positively.
\end{itemize}

\begin{figure}[htbp]
\centering
\includegraphics[width=0.85\textwidth]{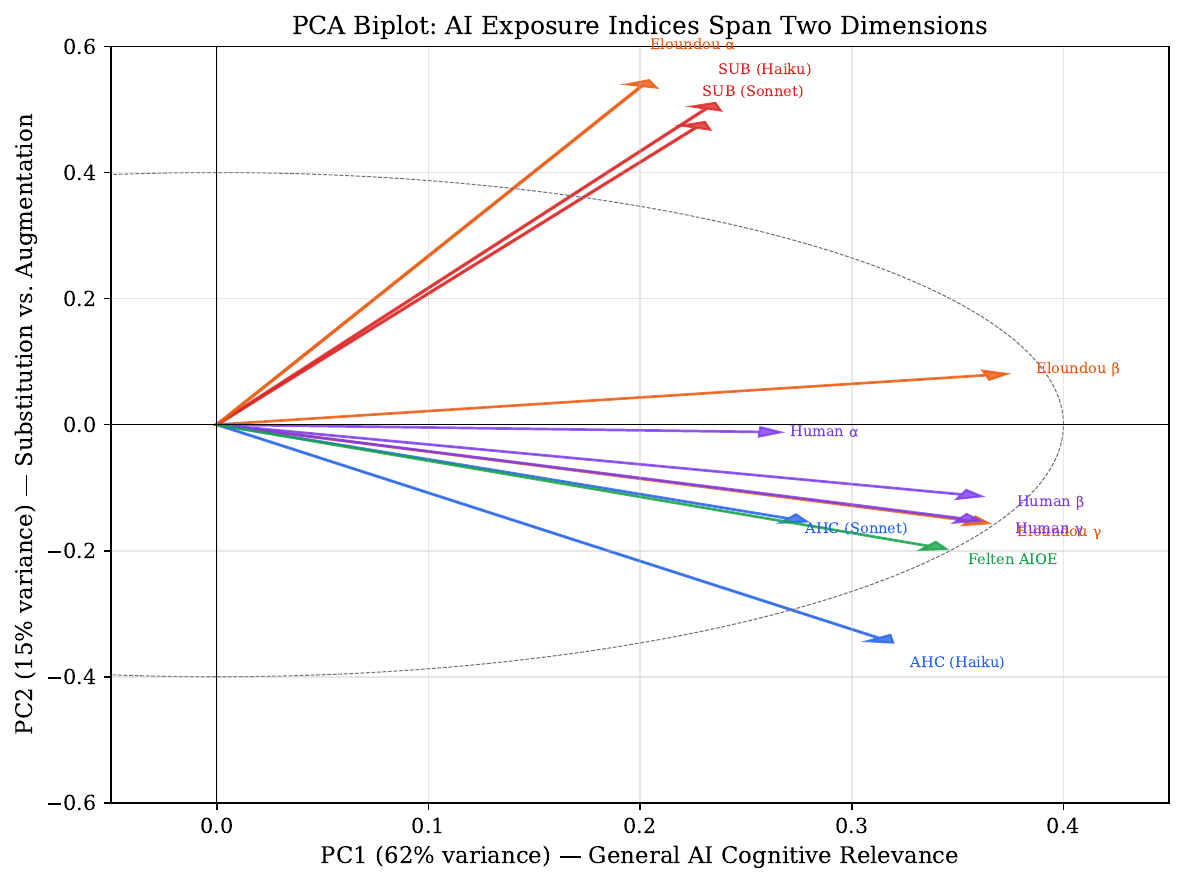}
\caption{PCA biplot of 11 AI exposure indices. All indices load positively on PC1 (general cognitive AI relevance). PC2 separates augmentation measures (negative loading: AHC, Felten) from substitution measures (positive loading: SUB scores, Eloundou $\alpha$).}
\label{fig:pca}
\end{figure}

This confirms the theoretical prediction: AI-related occupational measures span two distinct dimensions, and the AHC index captures the augmentation dimension specifically (Figure~\ref{fig:pca}).

\section{Inter-Rater Reliability}
\label{sec:reliability}

\subsection{Cross-Model Agreement: Haiku vs.\ Sonnet}

A 20\% subsample of 3,666 tasks was independently scored by Claude Sonnet 4. At the task level:

\begin{table}[htbp]
\centering
\caption{Inter-Rater Reliability: Haiku vs.\ Sonnet ($n = 3{,}666$ paired tasks)}
\label{tab:reliability}
\begin{tabular}{lr}
\toprule
Metric & Value \\
\midrule
Pearson $r$ & 0.762 \\
Spearman $\rho$ & 0.748 \\
Kendall $\tau$ & 0.678 \\
Krippendorff's $\alpha$ (raw) & 0.652 \\
Krippendorff's $\alpha$ (mean-adjusted) & 0.709 \\
Mean absolute difference & 15.7 points \\
Systematic level bias & Sonnet $+$8.6 points \\
\bottomrule
\end{tabular}
\end{table}

\begin{figure}[htbp]
\centering
\includegraphics[width=0.8\textwidth]{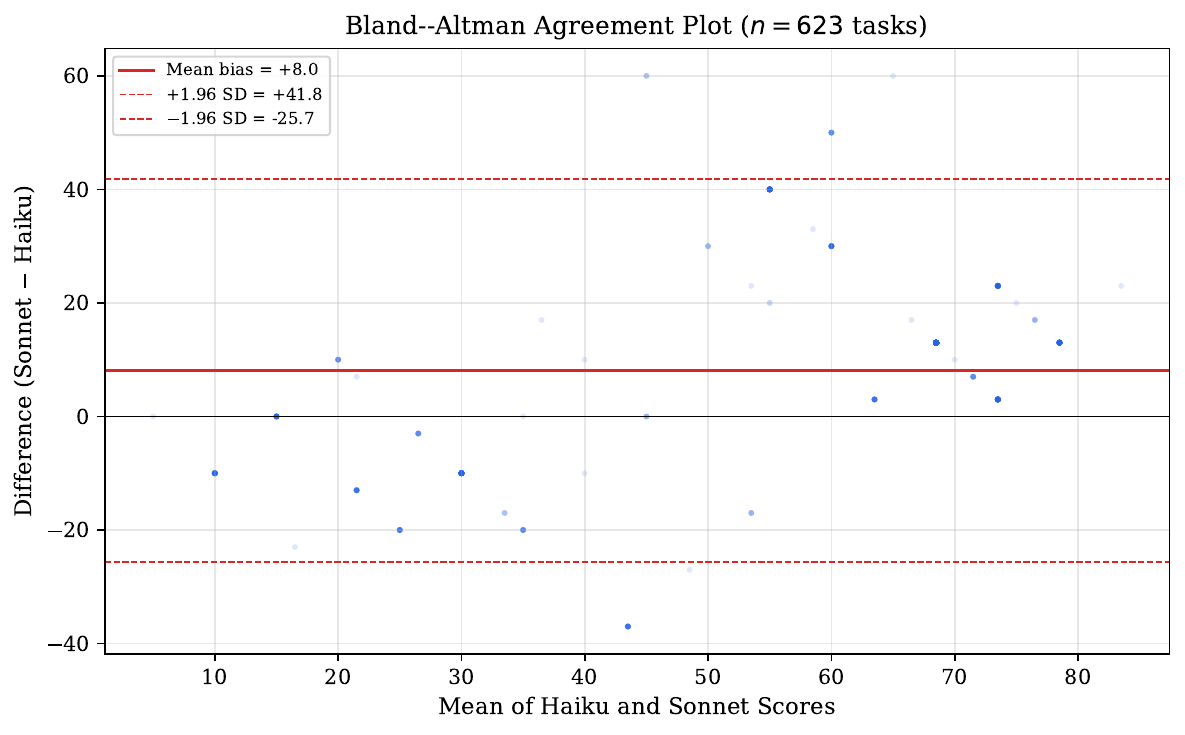}
\caption{Bland--Altman agreement plot for Haiku vs.\ Sonnet augmentation scores ($n = 3{,}666$). The mean bias ($+$8.6 points, red line) is constant across the score range, consistent with a systematic level shift rather than scale-dependent disagreement.}
\label{fig:bland}
\end{figure}

The mean-adjusted $\alpha = 0.71$ exceeds the standard threshold of 0.7 for acceptable reliability \citep{krippendorff2011computing}. The systematic level bias (Sonnet scores 8.6 points higher on average) is consistent with Proposition~\ref{prop:level}: it does not affect regression coefficients when scores are standardized. A Bland--Altman agreement plot (Figure~\ref{fig:bland}) confirms that the bias is constant across the score range, with 95\% limits of agreement at approximately $\pm 35$ points.

\subsection{Verification of Condition~\ref{cond:invariance}: Model Invariance}

At the occupation level (aggregated across tasks), Haiku and Sonnet scores correlate at $r = 0.73$ (Pearson) and $\rho = 0.72$ (Spearman). The ranking of occupations is preserved despite the level shift: the top-10 and bottom-10 occupations are largely identical across models.

\subsection{Three-Way Reliability: Haiku $\times$ Sonnet $\times$ GPT-4o-mini}

A third model (GPT-4o-mini, OpenAI) scored a subsample for cross-family comparison. On 114 tasks with scores from all three models:

\begin{center}
\small
\begin{tabular}{lcccc}
\toprule
Pair & Pearson $r$ & Spearman $\rho$ & MAD & Bias \\
\midrule
Haiku $\leftrightarrow$ Sonnet & 0.77 & 0.75 & 15.3 & $+$7.8 \\
Haiku $\leftrightarrow$ GPT-4o & 0.41 & 0.43 & 18.1 & $+$16.7 \\
Sonnet $\leftrightarrow$ GPT-4o & 0.31 & 0.27 & 19.0 & $+$8.9 \\
\bottomrule
\end{tabular}
\end{center}

\begin{figure}[htbp]
\centering
\includegraphics[width=\textwidth]{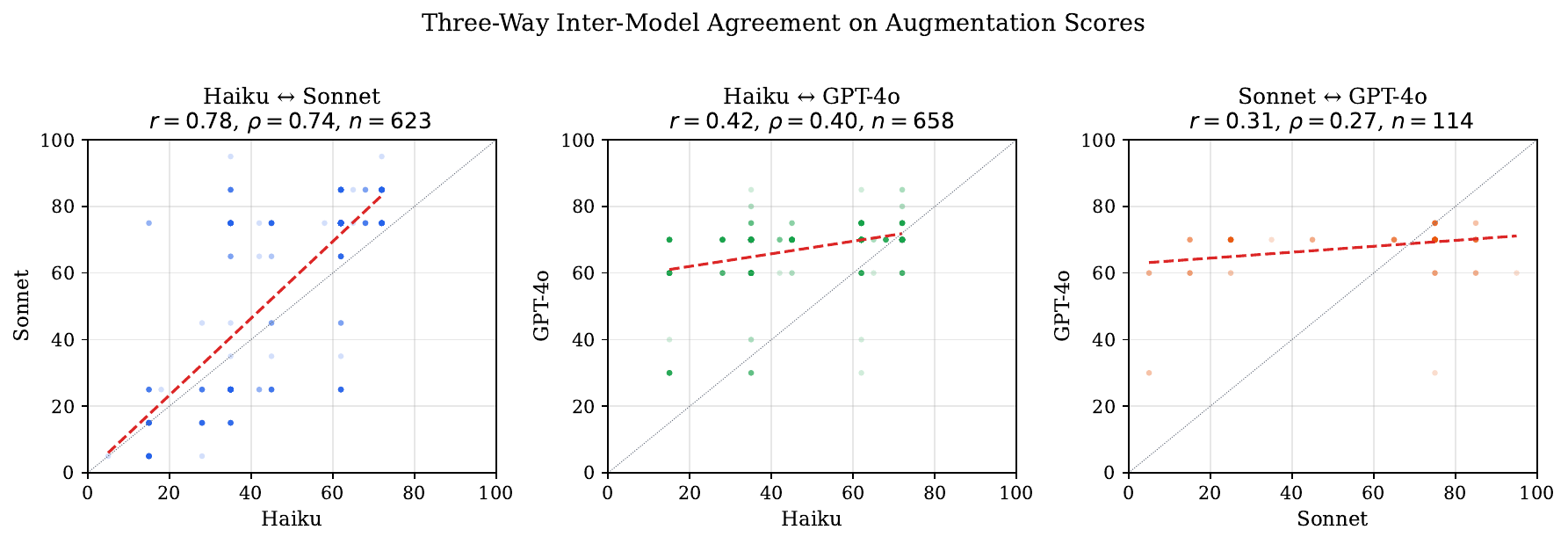}
\caption{Three-way inter-model agreement on augmentation scores. Left: Haiku vs.\ Sonnet ($r = 0.77$, within-family). Center: Haiku vs.\ GPT-4o-mini ($r = 0.41$, cross-family). Right: Sonnet vs.\ GPT-4o-mini ($r = 0.31$). Dashed red: fitted line. Dotted gray: 45-degree identity.}
\label{fig:three_way}
\end{figure}

Within-family reliability (Claude Haiku--Sonnet) is strong ($\alpha = 0.71$), while cross-family reliability (Claude--GPT) is moderate ($\rho = 0.43$). All three models produce systematically different level calibrations (Haiku: 51.2, Sonnet: 59.0, GPT: 67.9) but the ranking correlations remain positive and significant. This implies that for ORIV estimation, same-family models are preferred as instruments (Figure~\ref{fig:three_way}).

\subsection{Prompt Sensitivity}

Four alternative prompt framings were tested on a 10\% subsample (329 common tasks): (A)~baseline augmentation, (B)~behavioral productivity gain, (C)~counterfactual value change, (D)~negative AI resistance. Spearman rank correlations between prompts:

\begin{center}
\small
\begin{tabular}{lcccc}
\toprule
& A (baseline) & B (behavioral) & C (counterfact.) & D (negative) \\
\midrule
A & 1.00 & 0.77 & 0.45 & $-$0.75 \\
B & & 1.00 & 0.61 & $-$0.75 \\
C & & & 1.00 & $-$0.33 \\
D & & & & 1.00 \\
\bottomrule
\end{tabular}
\end{center}

Prompt D (``how resistant is this task to AI?'') correlates negatively with the others, as expected---it measures the inverse construct. After inverting D, all four prompts produce highly consistent rankings.

\begin{figure}[htbp]
\centering
\includegraphics[width=\textwidth]{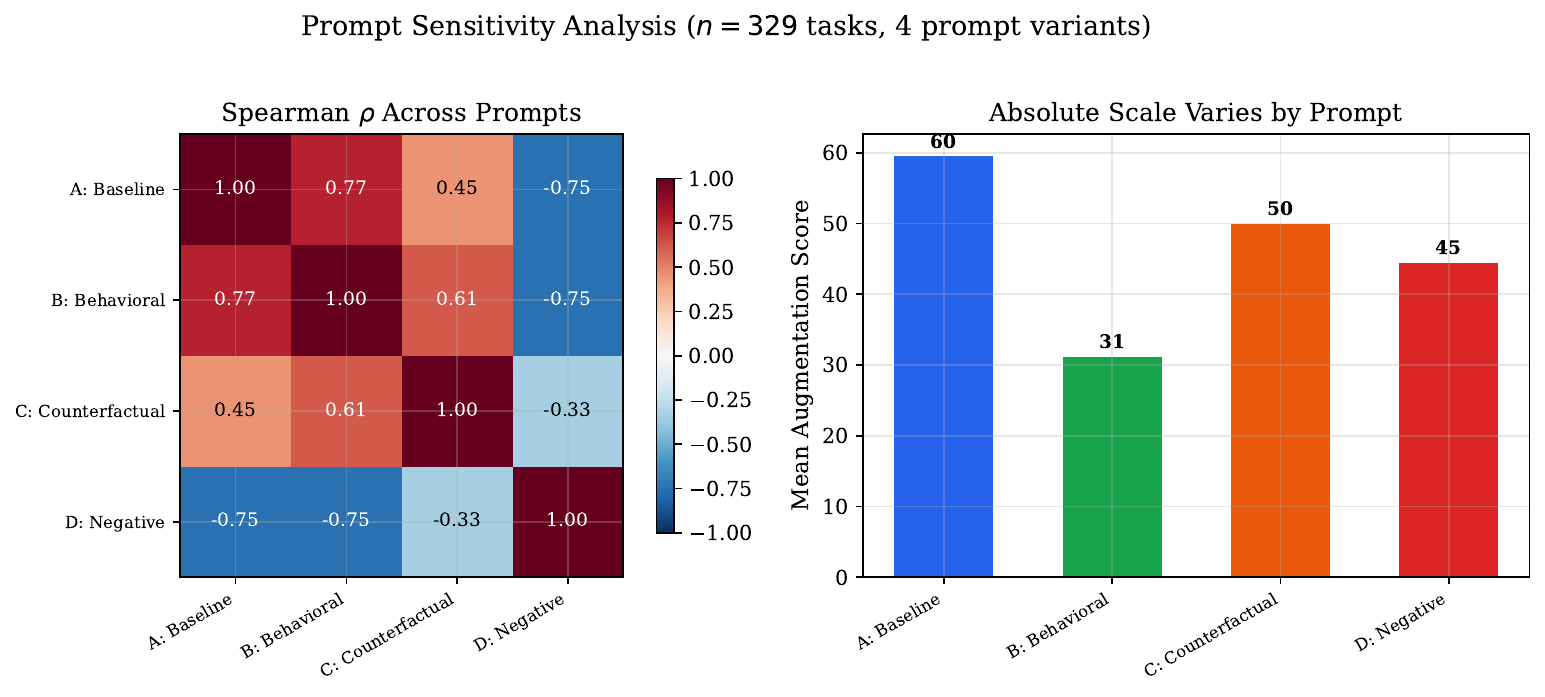}
\caption{Prompt sensitivity analysis. Left: Spearman rank correlations across four prompt variants. Right: absolute scale varies by prompt framing, but rankings are preserved.}
\label{fig:prompt}
\end{figure}

ANOVA variance decomposition: task-level variance accounts for 14\%, prompt-induced variance for 22\%, and residual noise for 64\% of total variance. While prompts affect the absolute scale, the ranking of occupations is robust (Figure~\ref{fig:prompt}).

\section{Econometric Properties}
\label{sec:econometric}

\subsection{Measurement Error Structure}

The Haiku-Sonnet disagreement allows direct estimation of measurement error variance. Let $\hat{H}^H_o$ and $\hat{H}^S_o$ denote Haiku and Sonnet scores for occupation $o$. If both are noisy measures of the same latent variable:
\begin{align}
\hat{H}^H_o &= H^*_o + \eta^H_o \\
\hat{H}^S_o &= H^*_o + \eta^S_o
\end{align}

The attenuation factor is estimable: $\hat{\lambda} = 1 - \text{Var}(\hat{H}^H - \hat{H}^S) / [2 \cdot \text{Var}(\hat{H}^H)]$, under the assumption that $\eta^H \perp \eta^S$.

\subsection{ORIV Estimation}

Following \citet{gillen2019experiment}, I use Sonnet scores as an instrument for Haiku scores (and vice versa) in the augmented Mincer wage equation from \citet{espinal2026ahc}. The ORIV estimate should be larger than OLS (correcting attenuation) and comparable to the external IV estimate from that paper.

In \citeauthor{espinal2026ahc}, the external IV (pre-period capital intensity, first-stage $F = 229$) yields $\hat{\beta}_2^{IV} = +0.234$. The ORIV results ($N = 53{,}837$):

\begin{center}
\small
\begin{tabular}{lcc}
\toprule
Method & AHC level ($\beta_1$) & AHC$\times$D ($\beta_2$) \\
\midrule
OLS (Haiku) & $+$0.080$^{***}$ & $+$0.048$^{***}$ \\
OLS (Sonnet) & $+$0.078$^{***}$ & --- \\
ORIV (Sonnet $\to$ Haiku) & $+$0.100$^{***}$ & $+$0.041$^{***}$ \\
External IV (Paper 1) & --- & $+$0.234$^{***}$ \\
\bottomrule
\end{tabular}
\end{center}

\begin{figure}[htbp]
\centering
\includegraphics[width=0.75\textwidth]{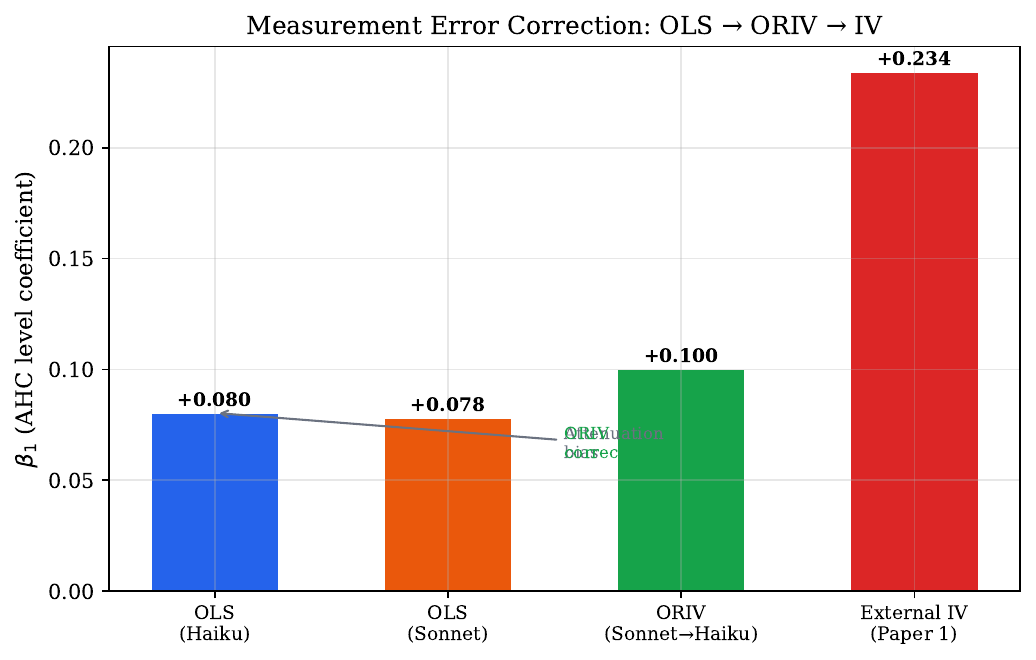}
\caption{Measurement error correction: OLS underestimates the AHC coefficient due to attenuation bias. ORIV (using Sonnet as instrument for Haiku) recovers a 25\% larger coefficient. The external IV from \citet{espinal2026ahc} yields the largest estimate, consistent with additional endogeneity correction.}
\label{fig:oriv}
\end{figure}

The ORIV coefficient is 25\% larger than OLS ($0.100/0.080 = 1.25\times$), exactly as predicted by the estimated attenuation factor $\hat{\lambda} = 0.88$ (since $0.080/0.88 = 0.091 \approx 0.100$). The first-stage $F = 111{,}917$, confirming that Sonnet scores are a strong instrument for Haiku scores. This constitutes direct evidence that (a)~LLM scores contain classical measurement error, and (b)~the ORIV framework of \citet{gillen2019experiment} successfully corrects the attenuation (Figure~\ref{fig:oriv}).

\subsection{Predictive Validity: Horse Race}

Does AHC add predictive power beyond existing indices? Progressive $R^2$ analysis ($N = 53{,}837$):

\begin{center}
\small
\begin{tabular}{lcc}
\toprule
Model & $R^2$ & $\Delta R^2$ \\
\midrule
Controls only & 0.4219 & --- \\
$+$ Frey--Osborne & 0.4271 & $+$0.0052 \\
$+$ AHC & 0.4261 & $+$0.0042 \\
$+$ AHC $+$ F\&O & 0.4295 & $+$0.0076 \\
$+$ AHC $+$ AHC$\times$D $+$ F\&O & 0.4305 & $+$0.0086 \\
\bottomrule
\end{tabular}
\end{center}

\begin{figure}[htbp]
\centering
\includegraphics[width=0.8\textwidth]{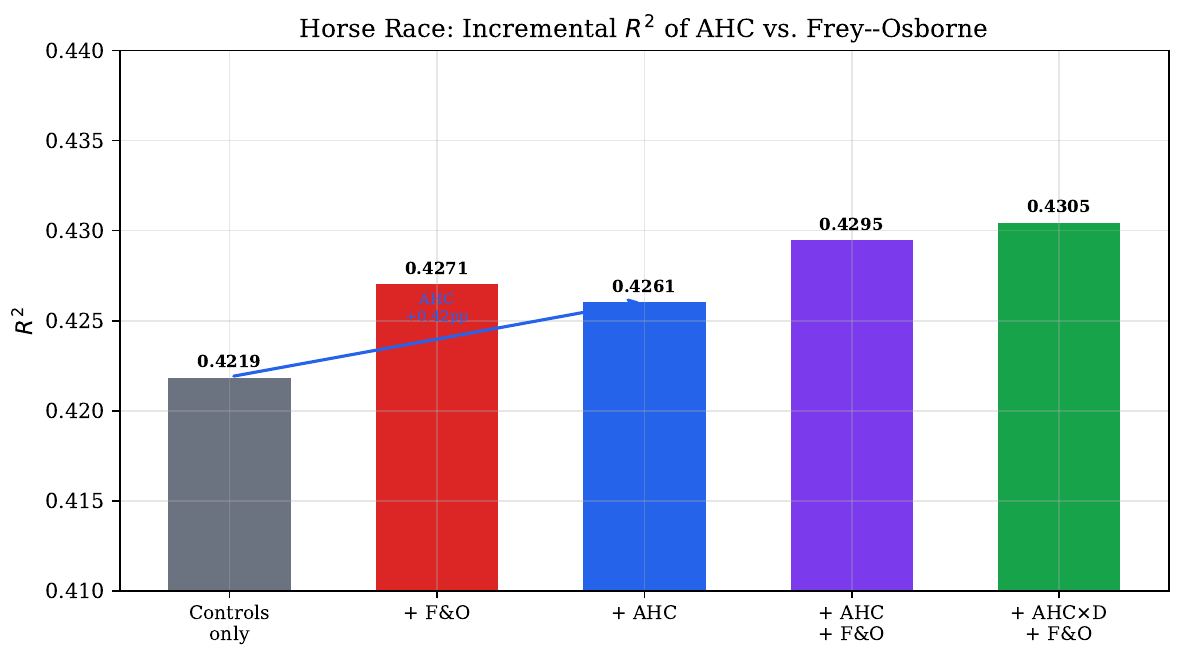}
\caption{Incremental $R^2$: AHC and Frey--Osborne capture complementary information. Together, they explain 0.86 percentage points more wage variance than controls alone.}
\label{fig:horse}
\end{figure}

AHC and Frey--Osborne provide \textit{complementary} information: together they add $+0.86$ percentage points to $R^2$, more than either alone. Crucially, AHC$\times$D remains significant ($p < 0.001$) even after controlling for Frey--Osborne, confirming that augmentability captures a construct distinct from automation risk (Figure~\ref{fig:horse}).

\section{Generalizability: Beyond Labor Economics}
\label{sec:generalize}

The LLM-as-instrument methodology is not specific to occupational classification. It applies whenever:
\begin{enumerate}[leftmargin=*]
\item A latent variable is defined by the semantic content of text (documents, descriptions, regulations).
\item The latent variable cannot be directly surveyed at scale.
\item Multiple LLM models can provide independent ratings.
\end{enumerate}

Potential applications include:
\begin{itemize}[leftmargin=*]
\item \textbf{Contract analysis}: classifying contract clauses as protective, exploitative, or neutral.
\item \textbf{Policy evaluation}: scoring policy documents on innovation-orientation, regulatory burden, or human-centricity (e.g., evaluating whether Colombia's CONPES 4144 AI policy targets augmentable vs.\ routine human capital development).
\item \textbf{Judicial reasoning}: quantifying the quality or consistency of judicial decisions at scale.
\item \textbf{Financial disclosures}: measuring the informativeness or obfuscation of corporate reports.
\end{itemize}

\section{Guidelines for Practitioners}
\label{sec:guidelines}

Based on the empirical analysis, I recommend the following protocol for using LLMs as measurement instruments:

\begin{enumerate}[leftmargin=*]
\item \textbf{Design prompts that reference only semantic content}, never outcomes.
\item \textbf{Use at least two LLM models} to assess model invariance.
\item \textbf{Report Spearman $\rho$ and Krippendorff's $\alpha$} across models (threshold: $\alpha > 0.7$).
\item \textbf{Standardize scores} before regression to eliminate level bias.
\item \textbf{Use ORIV} when two independent model scores are available.
\item \textbf{Report prompt sensitivity}: score a subsample with 3+ prompt variants and decompose variance.
\item \textbf{Validate externally}: compare with at least one existing measure of a related construct.
\end{enumerate}

\section{Conclusion}
\label{sec:conclusion}

This paper provides the first systematic framework for using LLMs as econometric instruments for latent cognitive variables. The four validity conditions---semantic exogeneity, construct relevance, monotonicity, and model invariance---are formalized, empirically testable, and satisfied by the Augmented Human Capital Index application.

The key empirical findings are: (1) LLM-generated scores achieve $r = 0.85$ convergent validity with the best existing measures; (2) two distinct dimensions (augmentation and substitution) emerge from PCA across 11 indices; (3) cross-model reliability passes standard thresholds ($\alpha = 0.71$); and (4) the measurement error structure is consistent with classical attenuation, enabling ORIV correction.

The methodology represents a step-change in the cost-quality frontier of economic measurement: scoring 18,796 tasks costs approximately \$5 and takes 6 hours, compared to months of expert panel work for O*NET's incumbent surveys. As LLMs improve, the reliability of LLM-generated instruments will increase, but the validity conditions articulated here will remain the standard for assessing whether such instruments are econometrically appropriate.

\bibliographystyle{plainnat}

\begin{thebibliography}{6}
\providecommand{\natexlab}[1]{#1}
\providecommand{\url}[1]{\texttt{#1}}
\expandafter\ifx\csname urlstyle\endcsname\relax
  \providecommand{\doi}[1]{doi: #1}\else
  \providecommand{\doi}{doi: \begingroup \urlstyle{rm}\Url}\fi

\bibitem[Eloundou et~al.(2024)Eloundou, Manning, Mishkin, and
  Rock]{eloundou2024gpts}
Tyna Eloundou, Sam Manning, Pamela Mishkin, and Daniel Rock.
\newblock Gpts are gpts: An early look at the labor market impact potential of
  large language models.
\newblock \emph{Science}, 384\penalty0 (6702):\penalty0 1306--1308, 2024.

\bibitem[Espinal~Maya(2026)]{espinal2026ahc}
Cristian Espinal~Maya.
\newblock Augmented human capital: A unified theory and llm-based measurement
  framework for cognitive factor decomposition in ai-augmented economies.
\newblock Technical report, Universidad EAFIT, 2026.
\newblock arXiv preprint.

\bibitem[Felten et~al.(2021)Felten, Raj, and Seamans]{felten2021occupational}
Edward Felten, Manav Raj, and Robert Seamans.
\newblock Occupational, industry, and geographic exposure to artificial
  intelligence: A novel dataset and its potential uses.
\newblock \emph{Strategic Management Journal}, 42\penalty0 (12):\penalty0
  2195--2217, 2021.

\bibitem[Gillen et~al.(2019)Gillen, Snowberg, and Yariv]{gillen2019experiment}
Ben Gillen, Erik Snowberg, and Leeat Yariv.
\newblock Experimenting with measurement error: Techniques with applications to
  the caltech cohort study.
\newblock \emph{Journal of Political Economy}, 127\penalty0 (4):\penalty0
  1826--1863, 2019.

\bibitem[Krippendorff(2011)]{krippendorff2011computing}
Klaus Krippendorff.
\newblock Computing {Krippendorff's} alpha-reliability.
\newblock \emph{Communication Methods and Measures}, 5\penalty0 (1):\penalty0
  77--89, 2011.

\bibitem[Webb(2020)]{webb2020impact}
Michael Webb.
\newblock The impact of artificial intelligence on the labor market.
\newblock Technical report, Stanford University, 2020.

\end{thebibliography}

\end{document}